\documentclass[12pt,preprint]{aastex}
\usepackage{times}

\newcommand{\VLSR}{V_{\rm LSR}}
\newcommand{\FeII}{[\ion{Fe}{2}]}

\newcommand{\kms}{km~s$^{-1}$}
\newcommand{\dotsec}{\rlap.^{''}}
\newcommand{\dotdeg}{\rlap.^{\circ}}

\shorttitle{\FeII\ JETS FROM L1551 IRS 5}
\shortauthors{Pyo, et al.}

\begin{document}
\title{SPATIO-KINEMATIC STRUCTURE AT THE BASE OF THE \FeII\ JETS FROM L1551 IRS~5\altaffilmark{1}}

\author{\sc Tae-Soo Pyo\altaffilmark{2}, Masahiko Hayashi\altaffilmark{2}, \\
Naoto Kobayashi\altaffilmark{4}, Hiroshi Terada\altaffilmark{2}, and Alan T. Tokunaga\altaffilmark{5} 
}
\altaffiltext{1}{Based on data collected at Subaru Telescope, which is operated by the National Astronomical Observatory of Japan.}
\altaffiltext{2}{Subaru Telescope, National Astronomical Observatory of Japan, 650 North A`oh\=ok\=u Place, Hilo, HI 96720, USA} 
\altaffiltext{3}{Institute of Astronomy, University of Tokyo, Mitaka, Tokyo 181-0015, Japan} 
\altaffiltext{4}{Institute for Astronomy, University of Hawaii, 2680 Woodlawn Drive, Honolulu, HI 96822, USA}

\email{pyo@subaru.naoj.org}

\begin{abstract}
We present observational results of the \FeII\ $\lambda$1.644 $\mu$m emission from the jets of L1551 IRS~5.  The data sets were obtained through 13 fully sampled slits aimed at the base of the jets. These sets are used to construct a three-dimensional cube.
The field of view was 5\rlap.{$''$}8$\times$4\rlap.{$''$}2.
We confirmed that the position of the knot PHK1 coincides with a stationary, point-like x-ray source within $\pm$0\farcs3. 
The northern and southern jets are distinguished from each other at a point 0\farcs6 away from their driving sources.
We also confirmed that the northern jet consists of well-separated high- and low-velocity components (HVC and LVC, respectively).
The HVC has a terminal velocity of $\sim$400~\kms\ and shows a consistently narrow velocity width of 40~\kms. 
The LVC covers the velocity range from $V_{\rm LSR}=\,$0 to $-$240~\kms\ and has broad velocity widths of $\sim$150--180 \kms. These decrease with distance from the driving sources.
The spatial width of the LVC varies from 0\farcs6--0\farcs7 at $\VLSR \sim$ $-$200~\kms\ to 0\farcs8--0\farcs9 at $\VLSR \sim$ $-$30~\kms. 
These characteristics are well understood in terms of the two types of outflow mechanisms that are working simultaneously: one is the HVC, which is launched in a narrow, inner radial region at 0.04--0.05~AU, and the other is the LVC, which is launched in a wider, outer radial region from within 0.1--4.5~AU of the accretion disk.
Part of the LVC emission could arise in the gas entrained or shocked by the HVC.
We also discuss the possibility that part of the HVC gas is thermalized at PHK1 to produce the x-ray emission and LVC.
\end{abstract}

\keywords{ISM: Herbig-Haro objects --- ISM: individual (HH 154, L1551) --- \\
\ \ \ \ \ \ \ \ ISM: jets and outflows --- stars:formation --- stars: pre-main-sequence}

\section{INTRODUCTION }\label{sec:intro}
L1551 IRS~5 is one of the best objects for the study of outflows and jets emanating from low-mass young stars. This is due largely to its proximity ($d \sim$ 140 pc), visibility, and isolation. 
It was the first object in which a molecular molecular outflow, with a CO radial velocity of $|V_{\rm r}| \la 20$~\kms, was discovered \citep{snell80}.
The CO outflow emission shows an apparent linear acceleration \citep{Uchida87,Msnell88}, which suggests that the outflow should consist of ambient material swept up by an invisible fast jet or wind.
A high-velocity ($|V_{\rm max}|~\sim~$100~km~s$^{-1}$) atomic wind was later detected inside the CO lobe \citep{Lizano88, Giovanardi92, Giovanardi00}. 
The existence of a fast wind can be deduced from the many optical and H$_2$ emission-line features showing large proper motions and radial velocities inside the blue-shifted lobe \citep{Stocke88,GH90,davis95}.
L1551 IRS~5 is also the first object toward which an optical jet was detected \citep{cohe82, mund83}. 
High angular resolution radio observations revealed that L1551 IRS~5 is a binary, or even triple, system of embedded young stellar objects, referred to as the L1551 IRS~5 VLA sources, with two jets emanating from them \citep{lim06, rodri03binary, rodri03jet, itoh00, hart00, fl98}. 

\begin{figure}
  \epsscale{0.7}
  \plotone{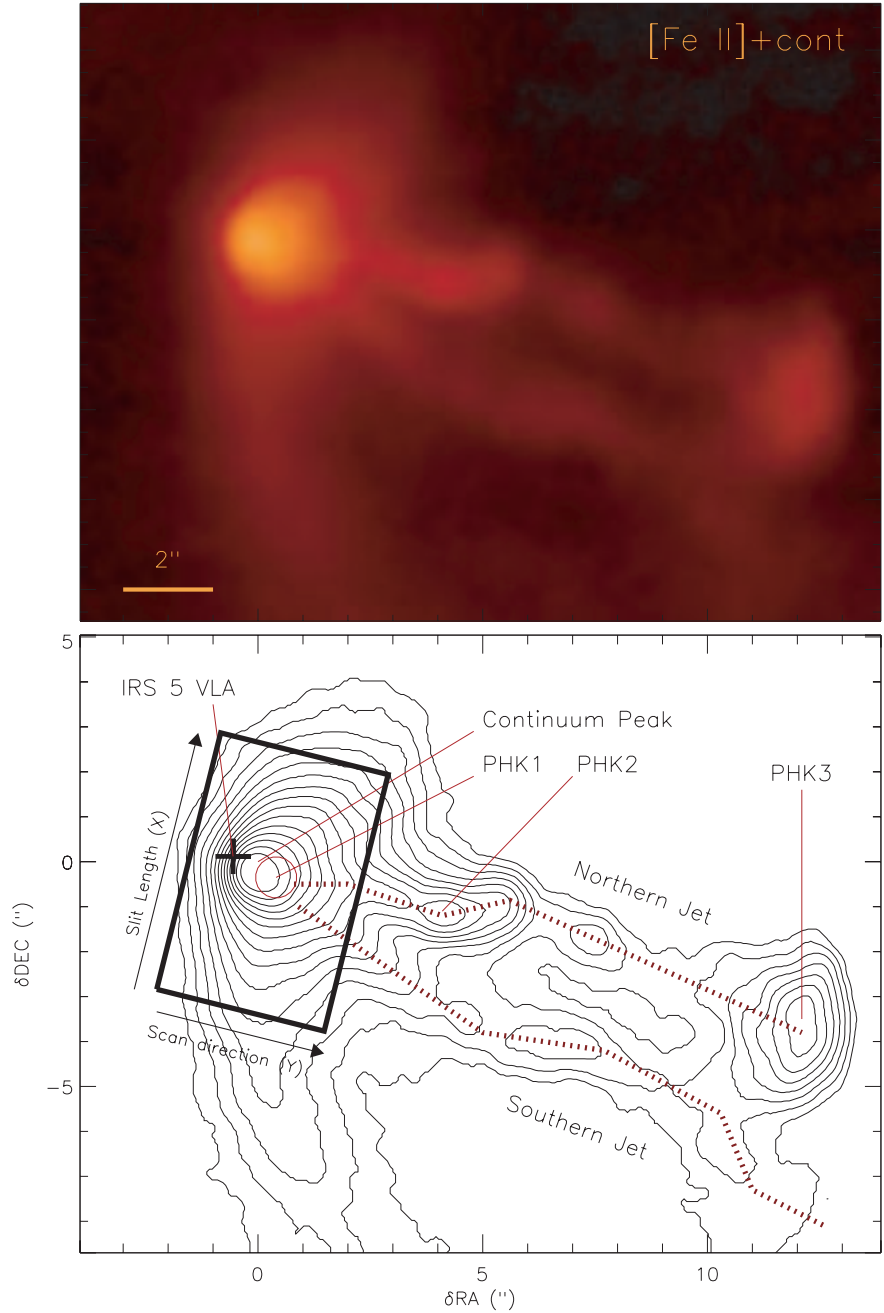}
  \caption{({\it upper}) An \FeII\ narrow-band filter (${\lambda}_c =$1.644 $\micron$, $\Delta \lambda =$0.026 $\micron$) image of the L1551 IRS 5 jets taken on 2002 November 27. 
Note that the emission includes both continuum and \FeII\ line.
({\it lower}) The area covered by the present slit-scan observations is shown with a thick solid-line rectangle superimposed on the contour version of the upper image.
The continuum peak, PHK1, PHK2 and PHK3 (Knot D) are indicated.
The red dotted-lines trace the brightest parts of the northern and southern jets.
The thick plus sign marks the mid point of the L1551 IRS~5 VLA sources \citep{camp88}.
The North is up and the East is left.
  \label{FeIIIm2002}}
\end{figure}

We show in Figure~\ref{FeIIIm2002} an image of the jets taken with an \FeII\ narrow-band filter on 2002 November 27 for reference.
It shows the two jets emanating from L1551 IRS~5 in the \FeII\ $\lambda$ 1.644 $\mu$m emission. 
The jets can be distinguished from each other at a position more than 3$\arcsec$ away from the driving sources in the figure.   
Along the northern jet, three strong knots (PHK1, PHK2, and PHK3) have been labeled \citep{pyo02}.
Several knots in the northern jet show proper motions along the flow, indicating episodic mass ejection from IRS~5 \citep{fridlund05,fl94,NS87}.
\citet{bachiller94} detected evidence of episodic mass loss in the molecular outflow along the axis of the blueshifted lobe.
The molecular outflow has a longer dynamical time scale (a few $\times$ 10$^4$ yr) than the optical jets (10$^{2}$--10$^{3}$ yr). 
The episodic mass-loss events could be related to the fact that IRS~5 is an FU~Ori type star \citep{carr87}.

We have been studying the kinematics and structure of jets at their bases by analyzing the near-infrared \FeII\ ($\lambda$1.644~$\mu$m) emission line which, due to its small extinction, traces jets closer to their driving sources than optical forbidden lines would.
This is the third of a series of papers on high spectral and spatial resolution \FeII\ spectroscopy of L1551 IRS~5 with {\it SUBARU}, an 8-meter-diameter ground-based optical and infrared telescope.
In previous papers \citep{pyo02,pyo05}, we explained that the northern jet has two distinct velocity components: the low-velocity component\footnote{The LVC in this paper should include the pedestal and red wing component (PWC), identified by \citet{pyo02, pyo05}, of the northern jet based on the results of this paper.} (LVC) with a velocity of $\sim-$100~\kms\ and the high-velocity component (HVC) with a velocity of $\sim-$300~\kms.
The LVC is much stronger than the HVC at PHK1, the brightest \FeII\ knot located close (1\farcs2 in 2002 December) to the L1551 IRS~5 VLA sources (hereafter called as IRS~5 VLA, a singular proper noun).
It shows a wide spatial width and a broad line width of 130--160~\kms\ near IRS~5 VLA, implying a large opening angle of 50--100$^\circ$. 
These characteristics are similar to the molecular Extremely High Velocity (EHV) outflows \citep{bachiller96}.
The HVC, on the other hand, consistently shows a narrow line width ($\sim$40~km~s$^{-1}$), is extended farther away from IRS~5 VLA, and is dominated in intensity at PHK2, a bright \FeII\ knot that was located $\sim$4\farcs7 away from IRS~5 VLA.
However, the previous observations did not fully cover the \FeII\ emission region, and the morphology and kinematics of the HVC and LVC, and their interrelation have remained unclear--especially at PHK1, where the northern and southern jets merge together spatially and the LVC is no longer separated from the southern jet because the two have a similar velocity.

The presence of two velocity components in jet emissions is an important clue to understanding the launching mechanism of outflows from young stellar objects.
Magnetocentrifugal acceleration has been widely accepted as a promising explanation for the launching mechanism of outflows from a young star-disk system \citep[see][]{konig00, Shu00, pudritz-ppv, Shang-ppv}.
There are two popular models based on this magnetocentrifugal mechanism: the ``$X$-wind model'' \citep{Shang-ppv} and the ``disk-wind model'' \citep{pudritz-ppv}. 
The $X$-wind model postulates that an outflow originates from a very narrow radial range of an accretion disk near its inner edge, where the highly concentrated stellar magnetic field interacts with the disk.
The disk-wind model assumes that an outflow is launched from the wider radial range of an accretion disk.
The terminal velocity of a magnetocentrifugally accelerated outflow scales with the Keplerian rotation velocity at its launching radius \citep{konig00}. 
Thus, the velocity width of a disk-wind should naturally be wider than that of an $X$-wind. 
This difference between the two models is best recognized in synthesized position-velocity diagrams (PVDs) \citep{pyo06}.
High spatial- and velocity-resolution spectroscopy is essential for the direct comparison of observations with these model calculations.

Here we present the results of \FeII\ line slit-scan observations covering the base of the L1551 IRS~5 jets with 13 fully sampled transverse slit locations.
We constructed a three-dimensional data cube to analyze the spatial-velocity structure of the \FeII\ jets.

\section{OBSERVATIONS AND DATA REDUCTION}\label{sec:obs}

Observations of the L1551 IRS~5 jets were conducted on 2004 December 1 (UT) using the Infrared Camera and Spectrograph \citep[IRCS,][]{koba00,toku98} mounted on the Cassegrain focus of the Subaru Telescope atop Mauna Kea, Hawai`i.
The {\it H}-band spectra were taken using the IRCS echelle spectrograph, which is equipped with a Raytheon 1024$\times$1024 InSb array with an Aladdin II multiplexer.
The near-infrared slit viewer of IRCS allowed us to place the slit accurately at selected positions on the jets. This was essential to match spectra with spatial coordinates.
We used a slit of 0\farcs6$\times$5\farcs9 (width$\times$length).
The slit width provided a velocity resolution of 60 \kms\ ($\lambda / \Delta\lambda =$5,000). 
The pixel scale along the slit length was 0\farcs060 pixel$^{-1}$.
The spectra were taken around PHK1 at 13 slit positions, at a position angle (PA) of 346$\arcdeg$, which was perpendicular to the northern jet. 
Adjacent slit positions were separated by 0$\dotsec$3, i.e. half of the slit width.
The field of view covered by the 13 slits was 5$\dotsec$9 $\times$ 4$\dotsec$2.
We made three sets of slit scans. Because of the variable sky conditions, only one set showed good data quality. 
That set was used for this paper. 
The total on-source exposure time was 180~s for each slit position. 
The seeing (FWHM) was $\sim$0\farcs4 at the {\it H}-band. 
The area covered by the current slit scan observations is shown in Figure~\ref{FeIIIm2002} (lower panel) by a thick rectangle, together with the coordinate system ($X$, $Y$) that we use in this paper. These are all superimposed on the contour versions of the upper panel image.
The positions of IRS~5 VLA, PHK1, PHK2 and PHK3 are also marked.

Sky spectra were obtained before and after each set of slit scans.
The atmospheric OH airglow emission was satisfactorily removed from the object spectra when the properly scaled sky spectra above the background level were subtracted from them.
In order to find a proper scaling factor for each object's spectrum when removing OH features, we used a strong OH line at $\lambda$1.6502~$\mu$m to derive a scaling factor so that the standard deviation of the residual (Object $-$ scaled Sky) at the line was minimized, while keeping the average of the residual at zero.
We observed the standard star HD~35656 (A0V, T$_{\rm eff} =\,$9480~K, $V=\,$6.403~mag) to use its spectrum for removing telluric lines and to calibrate flux.
The data were reduced mainly using the ECHELLE and LONGSLIT packages in IRAF \citep{pyo02}.  
After making a three-dimensional data cube with IRAF, we used IDL to analyze it.
 
\section{RESULTS}\label{sec:results}

\subsection{Continuum-Subtracted \FeII\ Image}\label{sec:fe2image}

\begin{figure} 
\epsscale{1.0}
\plotone{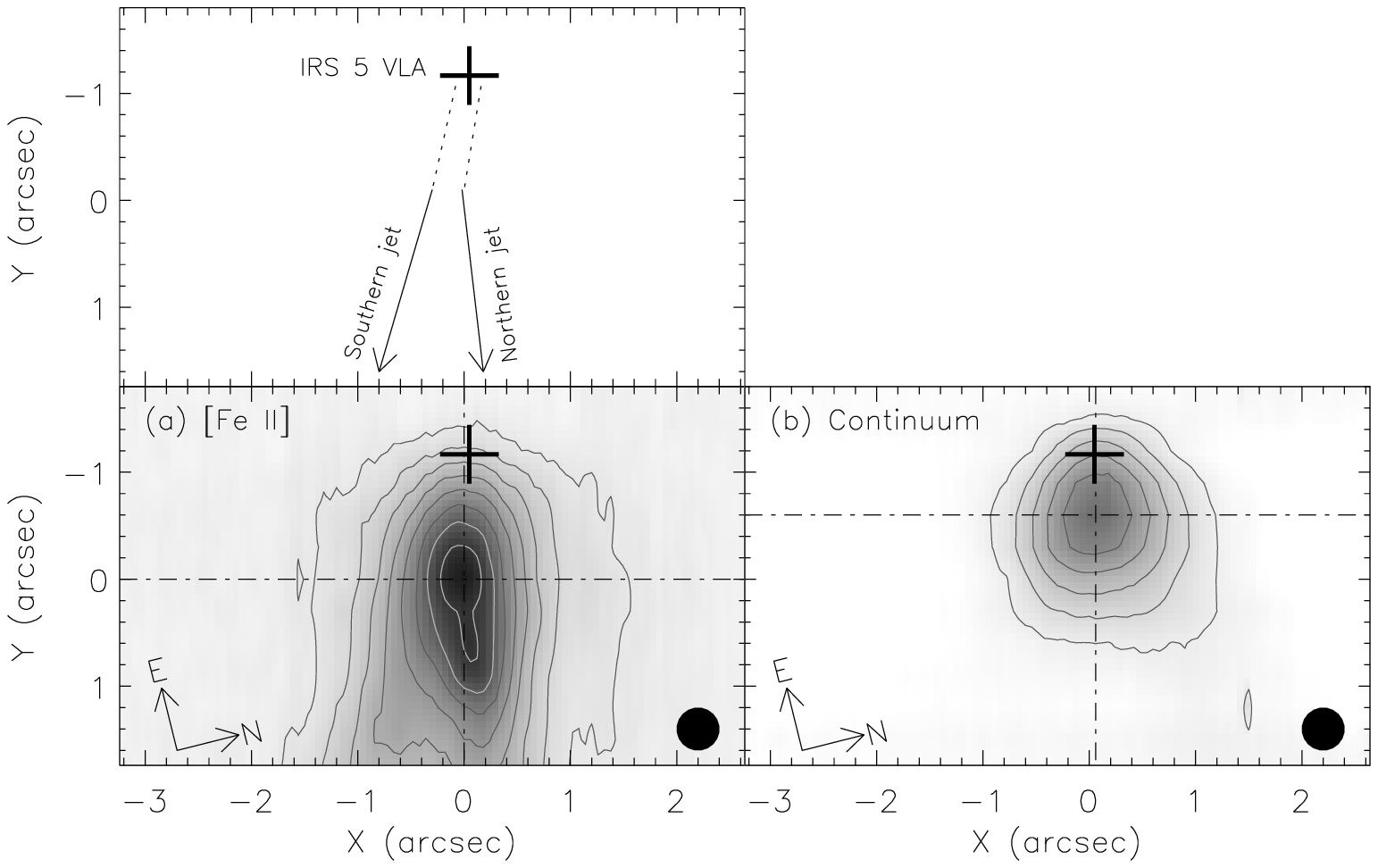}
\caption{
(a) Continuum-subtracted \FeII\ line image integrated from $-$420 to $+$20 \kms\ in $\VLSR$. 
The upper panel shows the directions of the northern and southern jets on $a$.
The dash-dotted horizontal and vertical lines define the peak of PHK1 at (0$\arcsec$, 0$\arcsec$). 
The thick plus sign marks the midpoint of the L1551 IRS 5 VLA sources \citep{camp88}. 
The contours are drawn with equal intervals in a logarithmic scale.
The contour levels are 0.12976, 0.30828, 0.51061, 0.73992, 0.99980, 1.29435, 1.62817, 2.00650, and 2.43529 $\times$ 10$^{-18}$ W m$^{-2}$. 
The filled circle at the lower right corner shows the seeing size of 0$\dotsec$4.
(b) Continuum emission averaged over the two line-free channels of $-$1840 to $-$560 \kms\ and $+$180 to $+$1500 \kms. 
The dash-dotted horizontal and vertical lines define the peak of the continuum at (0\farcs06, $-$0\farcs6).
Other signs and contour levels are the same as in $a$.
\label{FeII}}
\end{figure}

Figure~\ref{FeII}$a$ shows the continuum-subtracted \FeII\ line image integrated over the entire emission velocity of $-$420 to $+$20 \kms\ with respect to the local standard of rest.
Note that this is a pure \FeII\ line image in which the continuum emission is subtracted accurately down to the noise level. 
Slit-scan imaging allowed such accurate continuum subtraction because the continuum and \FeII\ line were observed at exactly the same time. 

It is difficult, on the other hand, to construct a pure line image from two narrow-band filter images ``on'' and ``off'' a given line emission if they are not taken simultaneously. This is because the continuous variation of sky conditions makes it difficult to match the continuum levels between the two images.

In Figure~\ref{FeII}$a$, the \FeII\ emission peak is called PHK1 (Paper~I) and is plotted at $X =$0$\arcsec$ and $Y =$ 0$\arcsec$, where $X$ is along the slit length (PA~$=$~346\arcdeg) and $Y$ is perpendicular to it (PA~$=$~256\arcdeg). 
A new knot is seen in Figure~\ref{FeII}$a$ at $Y =$ 0\farcs6 near the tip of the elongation.
The faint southern jet is seen at $X~\sim -$0\farcs6 and $Y \ga~$1$''$.
The northern and southern jets are merged and indistinguishable at PHK1.
The PAs of the northern and southern jets are 261$\dotdeg$6 $\pm$ 1$\dotdeg$2 and 235$\dotdeg$2 $\pm$ 1$\dotdeg$4, respectively.

Figure~\ref{FeII}$b$ shows the continuum emission averaged over the velocity ranges from $-$1840 to $-$560 \kms\ and from $+$180 to $+$1500 \kms\ with respect to the \FeII\ line rest wavelength in the LSR.
The continuum peak is located at ($X, Y$)  = (0\farcs06, $-$0\farcs6): the offset between the continuum peak and PHK1 is 0$\dotsec$6 ($\Delta\alpha = -$0$\dotsec$57, $\Delta\delta =-$0$\dotsec$20).

\subsection{Velocity Structure}\label{sec:v_structure}
\begin{figure}
\epsscale{1.0}
\plotone{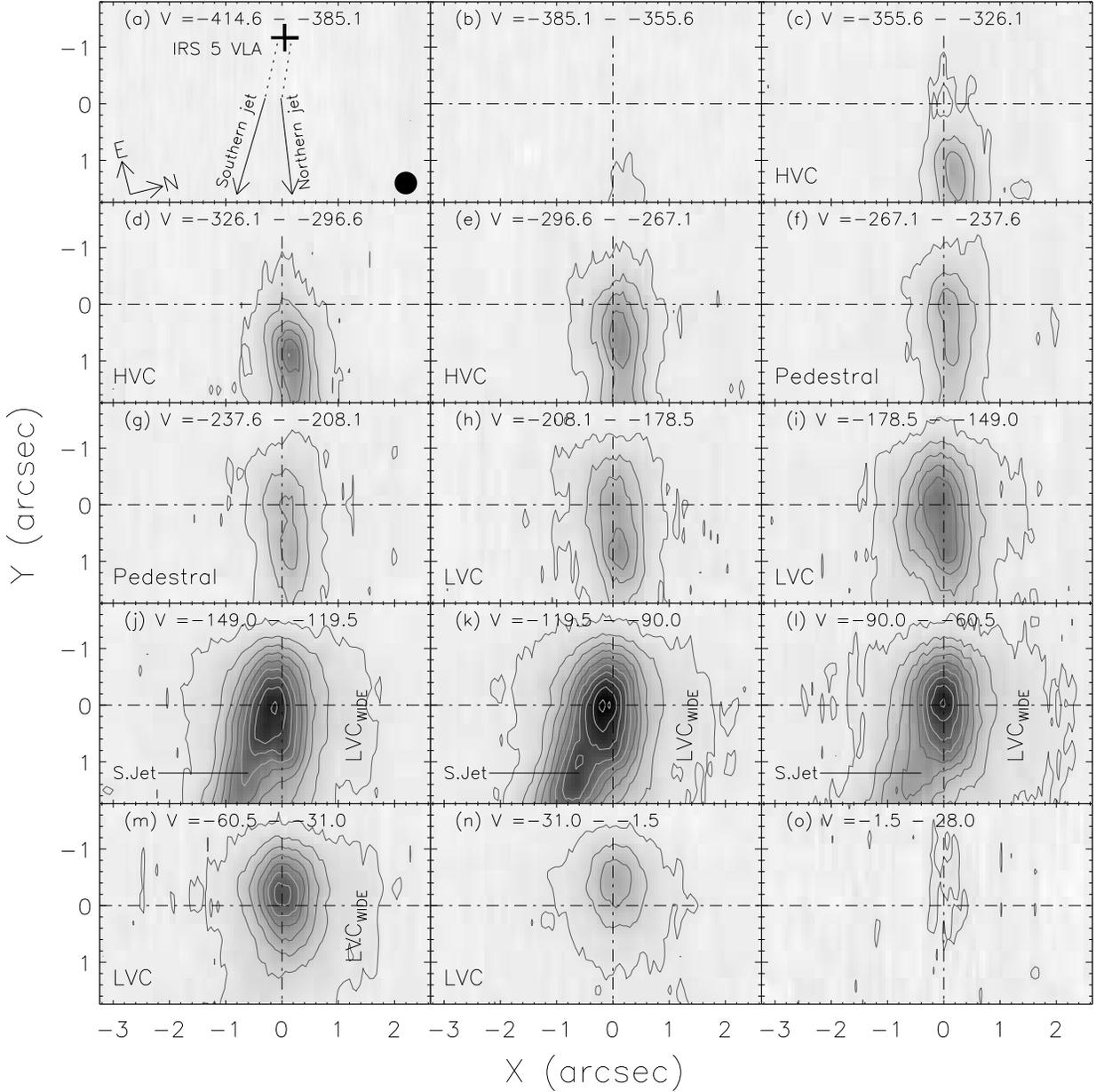}
\caption{
Channel maps of the \FeII\ $\lambda$ 1.644 $\micron$ emission. 
The velocity interval of each channel map is 29.5 \kms.
The contours are drawn with equal intervals in a logarithmic scale.
The contour levels are  0.01420, 0.05368, 0.09472, 0.13737, 0.18170, 0.22779, 0.27569, 0.32549, 0.37725, and 0.43106 $\times$ 10$^{-18}$ W m$^{-2}$.  
The filled circle at the lower right corner of panel $a$ shows the spatial resolution (seeing size) of 0$\dotsec$4.
\label{ChannelMaps}}
\end{figure}

Figure~\ref{ChannelMaps} shows the velocity channel maps of the \FeII\ $\lambda$1.644~$\micron$ emission for the velocity range from $V_{\rm LSR}=-$414.6 to $+$28.0 \kms. Each panel gives an integrated intensity map over a width of 29.5~\kms. 
The spatial width across the $X$-axis at the emission peak at each velocity channel is listed in Table~\ref{jet_width}.

\begin{figure}
\epsscale{1.0}
\plotone{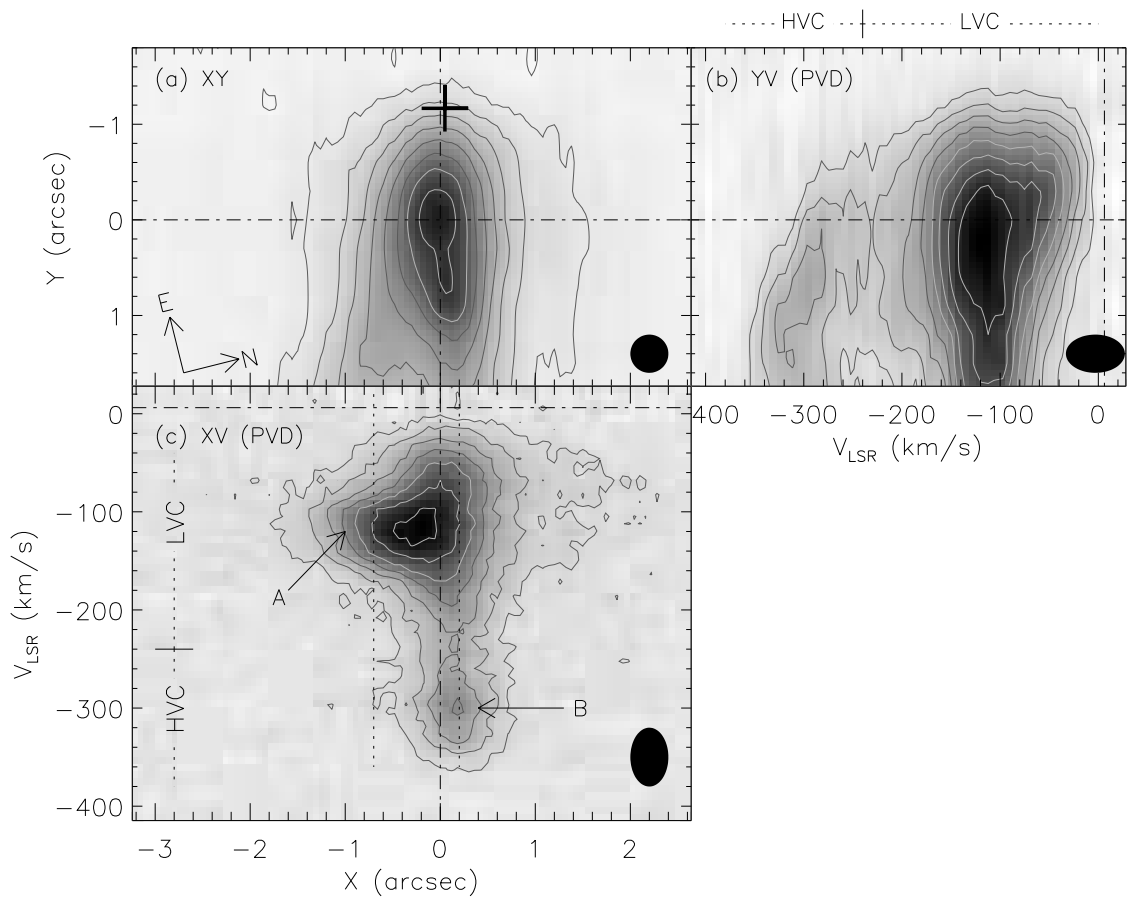}
\caption{
The data cube is shown from three points of view. 
(a) Same as Figure~\ref{FeII}$a$.
(b) The $Y$-velocity diagram integrated over the $X$-axis at each $Y$. 
The contour levels are 0.23560, 0.40585, 0.59604, 0.80863, 1.04626, 1.31187, 1.60877, and 1.94064 $\times$ 10$^{-18}$ W m$^{-2}$ \AA$^{-1}$.
(c) The $X$-velocity diagram integrated over the $Y$-axis at each $X$.
The contour levels are 0.23620, 0.55682, 0.93978, 1.39721, 1.94360, 2.59623, 3.37578, and 4.30693 $\times$ 10$^{-18}$ W m$^{-2}$ \AA$^{-1}$.
Label A shows the southern jet emission ($\VLSR = -$120 \kms) and Label B marks the high velocity component ($\VLSR = -$300 \kms) of the northern jet.
The contours are drawn with equal intervals in a logarithmic scale. 
The filled ellipse at the lower right corner of each panel indicates the resolutions of 0$\dotsec$4 in space and 60 \kms\ in velocity.
\label{IPVD}}
\end{figure}

The HVC is seen in the panels from \ref{ChannelMaps}$b$ to \ref{ChannelMaps}$e$ ($-$381.5 \kms\ $< V_{\rm LSR} < -$267.1 \kms), where the emission is elongated along the $Y$-axis with a spatial width of 0\farcs56--0\farcs68 (FWHM not deconvolved with the seeing size) centered around $X=~$0\farcs15 (see Table~\ref{jet_width}).
The emission peak moves from larger to smaller positions of $Y$, with the velocity in each panel reflecting the velocity gradient in the HVC. 
Although the channel maps in panels \ref{ChannelMaps}$f$ and \ref{ChannelMaps}$g$ ($-$267.1 \kms $< V_{\rm LSR} < -$208.1 \kms) show an elongated geometry similar to that of the HVC, both the HVC and LVC of the northern jet contribute to the emission in these channels (see Figure~\ref{IPVD}).
The emission in panel \ref{ChannelMaps}$h$ ($-$208.1 \kms $< V_{\rm LSR} < -$178.5 \kms) comes from the LVC without contamination from the southern jet.
The spatial width at this velocity is 0\farcs66, which is similar to that of the HVC. 
In the panels \ref{ChannelMaps}$i$--\ref{ChannelMaps}$l$ ($-$178.5 \kms $< V_{\rm LSR} < -$60.5 \kms), the southern jet overlaps with the LVC and the spatial width becomes as large as 1$''$.
Channel maps \ref{ChannelMaps}$m$ and \ref{ChannelMaps}$n$ ($-$60.5 \kms $< V_{\rm LSR} < -$1.5 \kms) are free from the southern jet emission and show the less-blueshifted, lower-velocity emission from the northern jet.
The spatial widths at these velocities are 0\farcs86 and 0\farcs83, respectively, and are significantly larger than those of the LVC in panels \ref{ChannelMaps}$g$ and \ref{ChannelMaps}$h$, which show more-blueshifted, higher-velocity emission from the northern jet.

Figure~\ref{IPVD}$b$ shows a position-velocity diagram (PVD) where the \FeII\ flux integrated over the  $X$-axis (perpendicular to the northern jet) at each position of $Y$ is plotted against the radial velocity ($V_{\rm LSR}$). 
This shows the well-separated high- and low-velocity components discussed above: the HVC at $V_{\rm LSR}\sim-$300~\kms\ and the low-LVC covering $V_{\rm LSR}$ from $-$240~\kms\ to 0~\kms.

Figure~\ref{IPVD}$c$ shows an $X$-velocity diagram (XVD) where the \FeII\ flux integrated over the $Y$-axis (parallel to the northern jet) at each position of $X$ is plotted against the radial velocity. 
An emission feature is seen at $\sim-$120~\kms\ corresponding to the southern jet (marked as ``A'') in this figure.
The HVC is also seen at $\sim-$300~\kms\ around $X\sim\,$0\farcs2 (marked as ``B'').   

\begin{figure}
\epsscale{1}
\plotone{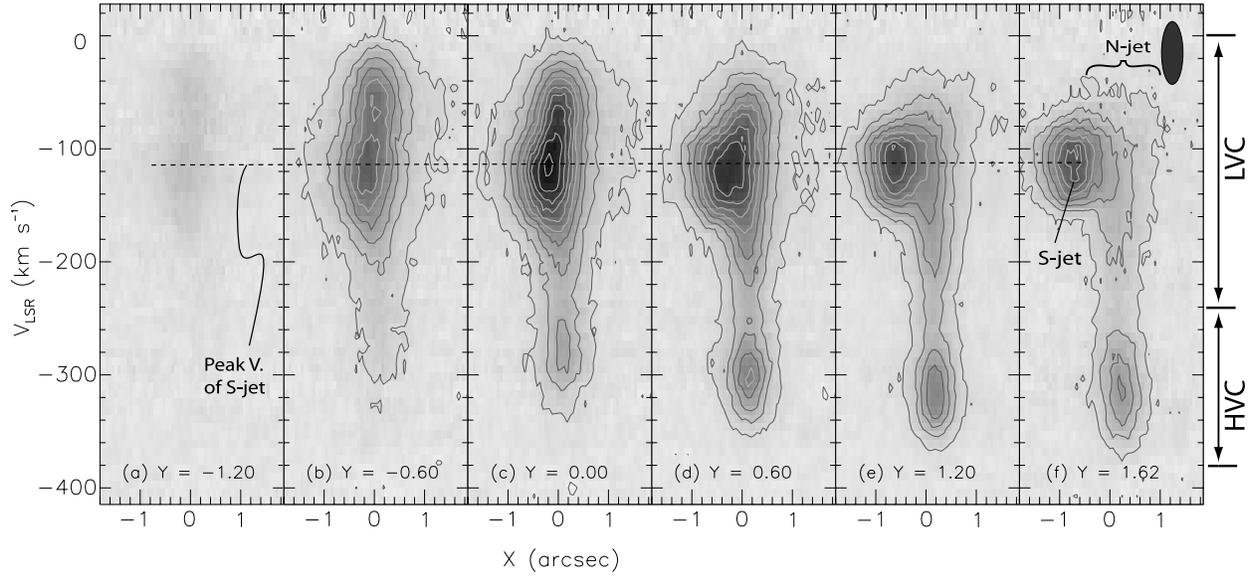}
\caption{$X$-velocity diagrams at $Y =$ (a) $-$1$\dotsec$2, (b) $-$0$\dotsec$6, (c) 0$\dotsec$0, (d) 0$\dotsec$6, (e) 1$\dotsec$2, and (f) 1$\dotsec$62.
The contour levels are 0.00627, 0.01667, 0.02717, 0.03778, 0.04850, 0.05932, 0.07026, 0.08131, 0.09247, and 0.10375 $\times$ 10$^{-18}$ W m$^{-2}$ \AA$^{-1}$.
The filled ellipse at the upper right corner of panel $f$ shows the resolutions of 0$\dotsec$4 in space and 60 \kms\ in velocity.
The dotted horizontal line traces the peak velocity of the southern jet. 
\label{VX_PVDs}}
\end{figure}

Figure~\ref{IPVD}$c$ can be deconstructed along the $Y$-axis to produce Figure~\ref{VX_PVDs}, a series of XVDs at various positions of $Y$.
This is one of the most remarkable diagrams created from the slit scan observational data. 
First, the southern jet can be traced down to $Y$=$-$0\farcs60, i.e. as close as 0\farcs6 from the driving source IRS~5 VLA.
This is because the radial velocity of the southern jet is almost constant at $\VLSR \sim$ $-$120$\pm$10 \kms\, with a relatively narrow FWHM width of 80.5$\pm$15.3 \kms\ (FWHM)\footnote{The FWHM width is 53 \kms\ if deconvolved with the velocity resolution of 60 \kms}, in contrast to the large velocity width of the LVC (150--180~\kms\ FWHM).

Even at $Y\la$0\farcs6 where the LVC mingles with the southern jet, the southern jet can be separated from the LVC.
Note that the southern jet can be traced only at $Y \ga$1$\arcsec$ ($\ga$2$\arcsec$ from the driving source) in the velocity-integrated spatial map in Figure~\ref{FeII}$a$ (or Figure~\ref{IPVD}$a$).
Figure~\ref{VX_PVDs} shows that the southern jet gradually drifts away from $X =-$0\farcs1 to $-$0\farcs8 as $Y$ increases from $-$0\farcs60 to 1\farcs62.
The flux density of the southern jet reaches its maximum at $Y$=0$''$, i.e. at PHK1, where it is 2.5 times larger than that at $Y$=1\farcs62.
This means that the southern jet significantly contributes to the \FeII\ flux of PHK1.

Second, the lower (less-blueshifted side) velocity cutoff of the LVC tends to decrease with $Y$.
This is seen in Figure~\ref{VX_PVDs} as the lowest contours intersect the $X$=~0 line at $\VLSR \sim$ 0 \kms\ for $Y\la~$0$''$; this occurs at $\sim-$50 \kms\ for $Y\ga~$1\farcs2.
The higher (more-blueshifted side) velocity cutoff of the LVC is rather constant in this range of $Y$.
An interesting point is that the decrease in the lower velocity cutoff of the LVC is similar to that seen in the HVC: its peak velocity decreases from $\VLSR = -$263 to $-$312 \kms\ with increasing $Y$ from 0$''$ to 1$''$.
However, we conclude that the two similar tendencies are unrelated because the velocity gradient of HVC seen for 0$''<Y<$1$''$ is time variable (see \S\ref{sec:comparison}), while the lower cutoff velocity gradient of LVC does not vary with time, as seen in the PVD taken four years ago.

\subsection{Comparison with Previous Data}\label{sec:comparison}
In this section we summarize some of the important points for the discussion below by comparing the current \FeII\ data with previous spectroscopic data taken in 2000 December (Paper~I) and 2002 November (Paper~II).
\begin{enumerate}  
\item{The pure \FeII\ line image (Figure~\ref{FeII}$a$) bears a close resemblance to Figure~2$a$ (Paper~I) that was constructed from two narrow-band filter images ``on'' and ``off'' of the \FeII\ line.
This means that the continuum was satisfactorily subtracted in the previous work.
Because of the slightly improved seeing, or possible time variation between 2000 and 2004, Figure~\ref{FeII}$a$ shows the elongation of PHK1 along the northern jet more clearly than Figure~2$a$ in  Paper~I.}

\item{The offset between the {\it H}-band continuum peak and PHK1 remains the same as before. 
It was 0\farcs56 in 2000 December (Paper I)\footnote{Note that in \citet{pyo02} the separation between the continuum peak and PHK1 was reported to be 0\farcs7, which was measured using the incorrect pixel scale of 0\farcs075~pixel$^{-1}$.  The actual pixel scale was 80\% of this value and the true separation was 0\farcs56 \citep[see][]{pyo03}.} and 0\farcs60 in 2004 December (present work).
The difference in the offsets is within the measurement error of $\sim$0\farcs1 and is much smaller than that expected from the proper motions of 0$\dotsec$2--0$\dotsec$44 yr$^{-1}$ for the other knots \citep{fridlund05,fl94,NS87}.
Because the continuum peak position seems stationary, or at least its proper motion is much smaller than those of the other knots, PHK1 may also be stationary with respect to the IRS~5 VLA. 
Previous measurements (noted in Paper~I) revealed that, within an uncertainty of 0\farcs25, the {\it H}-band continuum peak coincides with the {\it K}-band peak position, which is located at 0\farcs57 $\pm$ 0\farcs35 ($\Delta\alpha =-$0\farcs55, $\Delta\delta =-$0\farcs13) away from IRS~5 VLA \citep{camp88}.
This places IRS~5 VLA at $Y =-$1\farcs2 $\pm$ 0\farcs4.}

\item{The $Y$-velocity diagram (YVD) in Figure~\ref{IPVD}$b$ is compared with Fig.~2$b$ in Paper~I, which is a YVD taken in 2000 December, but is not integrated over $X$ at each position of $Y$. 
Although both YVDs show the two velocity components, there is a notable difference in the radial velocities of the HVC.
The radial velocity of HVC does not vary largely with $Y$ for 0$''<Y<$1$''$ in the 2000 December data (Figure~2$b$ in Paper~I), while it decreases from $-$263 to $-$312 \kms\ with $Y$ for 0$''<Y<$1$''$ in the 2004 December data (Figure~\ref{IPVD}$b$).
This difference is a result of time variation because the radial velocity gradient of HVC seen in Paper~I is not reproducible from any cross-section of the current data cube (see Figure~\ref{ChannelMaps}).}

\item{Figure~\ref{IPVD}$c$ (XVD) is compared with Fig.~1$a$ of Paper~II, which is an XVD taken in 2002 November but is not integrated over $Y$.
The two diagrams share common features, except that the southern jet at $V_{\rm LSR} \sim -$ 120 \kms\ for $X<$0$''$ is more conspicuous in Figure~\ref{IPVD}$c$ because it is integrated over $Y$ where the southern jet gives significant contribution.
The lowest contours in these figures show a very large spatial width corresponding to the spatially wide subcomponent of LVC (LVC$_{\rm WIDE}$) reported in Paper~II.}

\item{Figure~\ref{VX_PVDs} suggests that the broad, ``flat topped'' \FeII\ line profiles in LVC that were reported in Paper~I at $Y=-$0\farcs4 to $-$0\farcs8 are most likely formed as a result of the overlapping emission from the LVC and the southern jet.}

\item{Figure~\ref{VX_PVDs} shows that the less-blueshifted side cutoff velocity of the LVC decreases (becomes more blueshifted) with $Y$.
As a result, the velocity width of the LVC decreases with distance from IRS~5 VLA.
It is also clear from Figure~\ref{VX_PVDs} that this tendency is not significantly affected by the southern jet emission.
Thus, the velocity narrowing of the LVC along the outflow, which was pointed out in Paper~I (Figure~2$b$ there), may be related only to the northern jet.
We interpreted this tendency as due to the collimation of the LVC in Paper~I, and that interpretation remains valid.}
\end{enumerate}

\section{DISCUSSION}\label{sec:discussion}
\subsection{The Two Velocity Components: HVC and LVC}\label{sec:two_vc}
The present slit scan observations provided unprecedented spatio-velocity information for the two jets emanating from IRS~5 VLA at the spatial and velocity resolutions of 0\farcs4 and 60 \kms, respectively.
This has enabled us to separate the northern and southern jets even in the close vicinity ($<$1$''$) of IRS~5 VLA (Figure~\ref{VX_PVDs}), although the two jets are discernible only at $\ga$2$''$ away from IRS~5 VLA if no velocity information is available.

There is no clear evidence to show which of the two IRS~5 VLA sources is responsible for the northern or southern jets.
The northern and southern radio jets detected within $\sim$1$''$ of IRS~5 VLA \citep{rodri03jet} have PAs of 247\arcdeg and 235\arcdeg respectively, which are similar to those of the northern (262\arcdeg) and southern (235\arcdeg) \FeII\ jets.
Although the PA of the northern radio jet is somewhat smaller than that of the \FeII\ jet, it bends toward the west at 0\farcs7 west-southwest of the northern driving source and its PA changes to $\sim$260\arcdeg, which agrees well with that of the northern \FeII\ jet. 
If we assume that the northern and southern radio jets are smoothly connected to their respective \FeII\ counterparts at $\sim$1$''$ west of IRS~5 VLA, it is natural that the northern and southern radio jets are the inner parts of northern and southern \FeII\ jets, respectively.

While the southern jet shows a single velocity component with a relatively narrow line width and little radial velocity variation, the spatial and velocity structure of the northern jet is more complicated.
Below, we summarize the characteristics of the northern jet revealed by our current and previous observations of the \FeII\ $\lambda$1.644~$\mu$m line.
\begin{enumerate}
\item{The northern jet covers a wide velocity range of $-$380 to 0 \kms.}
\item{The position-velocity diagrams along and across the northern jet show two distinct components with different radial velocities: one is the HVC at a velocity of $\sim-$300 \kms\ and the other is the LVC covering the velocity range of $-$240 to 0 \kms.}
\item{The HVC shows a narrow line width with a deconvolved FWHM width of $\sim$40 \kms\ (see Paper~I).}
\item{The LVC shows a broad line width of 130--160 \kms\ (FWHM, see Paper~I). 
This width is intrinsic to the LVC of the northern jet and is not caused by the contamination from the southern jet (see item 6 of \S\ref{sec:comparison}).}
\item{The LVC shows a spatial width of $\sim$0\farcs65 at its more-blueshifted velocity side ($V_{\rm LSR}=-$200 \kms). 
This width is similar to that of the HVC.  
The LVC has a larger spatial width ($\sim$0\farcs85) at its less-blueshifted velocity side ($V_{\rm LSR}=-$30 \kms).}
\end{enumerate}

An important issue to consider is whether the HVC and LVC of the northern jet represent two types of jets from a single driving source or merely an inner and an outer part of a single jet.
We discussed this in Papers~I and II, with preference to the presence of two outflows from a single driving source. This is because the HVC and LVC are identified as two distinct peaks in the PVDs and show different kinematical characteristics as typically seen in their velocity widths.
The HVC and LVC are, on the other hand, not easily discernible in the channel maps of Figure~\ref{ChannelMaps}, which might suggest that the HVC and LVC are smoothly connected.
However, we should note that it is difficult to theoretically simulate PVDs with two distinct peaks under the assumption of monotonically increasing or decreasing variation of physical conditions from the jet axis to its outer annuli.
The two distinct peaks in the PVDs may thus imply that there are two different radial regions flowing at different velocities in the northern jet, if axial symmetry holds, and that the two annuli emit equally strong \FeII\ emission, even if they are part of a single outflow.

Another point that has to be stressed is that the HVC always shows a narrow line width, while the LVC has a larger line width for the four \FeII\ jets so far observed at sufficiently high spatial and spectroscopic resolutions.
Table~\ref{velocity_width} summarizes the deconvolved velocity widths (FWHM) of the HVCs and LVCs for those \FeII\ jets.
The HVCs of the jets from DG~Tau, RW~Aur, and HL~Tau are conspicuous and are easily identified in their PVDs. They all show a width of $\sim$50 \kms\ with their blueshifted radial velocities of $\sim$200 \kms.
The LVC of the DG~Tau jet is extended and its broad velocity width is clearly seen in Figures~1 and 2 of \citet{pyo03}.
The LVCs of RW~Aur and HL~Tau are compact and are detected as less-blueshifted wing or shoulder emission at or near the positions of their driving sources.
These two LVCs do not show distinct peaks in their PVDs, although the L1551 IRS~5 northern and DG~Tau jets do.
The LVC emission from all these sources fills the radial velocity range between the HVC and $\sim$0~\kms, suggesting a velocity width of 100~\kms\ or larger.
The southern jet of L1551 IRS~5 is, in this sense, categorized as an HVC. This is based on its typical deconvolved velocity width of 53 \kms\ and extended geometry, and is as listed in Table~\ref{velocity_width}.
The current paradigm of magnetocentrifugally accelerated outflow models predicts that the terminal velocity of a jet is mainly determined by the Alfv\'en radius, within which the magnetic field lines work as lever arms \citep{konig00,Shu00,shang98,kudoh98}.
Thus, the narrow velocity width of a jet indicates that the line-emitting gas is launched within a narrow radial range of an accretion disk, while the broad velocity width of a jet (after it becomes collimated) implies a wider range of launching radius.  
On the basis of this assumption, \citet{pyo06} suggested that the HVCs of RW~Aur and HL~Tau should be launched from a narrow range of radius or that only such outflowing gas originating from a narrow radial range should be sufficiently excited to contribute to the line emission.
The same interpretation can be applied for the HVCs from L1551 IRS~5 and DG~Tau, judging by the similarities of their HVCs to those of RW~Aur and HL~Tau.

On the other hand, the LVCs show a velocity width 2--4 times larger than the HVCs (Table~\ref{velocity_width}), meaning that the LVC-emitting gas is launched from the wider outer radial range of an accretion disk if the streamlines are collimated; i.e. the LVC is extended more than a few tens of AU from their driving sources.
Because this length scale corresponds to $\sim$0\farcs2 at the assumed distance of 140~pc, the spatially extended LVCs observed for the northern jet of L1551 IRS~5 and the DG~Tau jet are consistent with the outflow launched from a larger radial range of an accretion disk. There is also a possibility that the spatially compact LVCs toward RW~Aur and HL~Tau may have arisen in a region where the streamlines are not yet collimated.

\begin{figure}
\epsscale{1}
\plotone{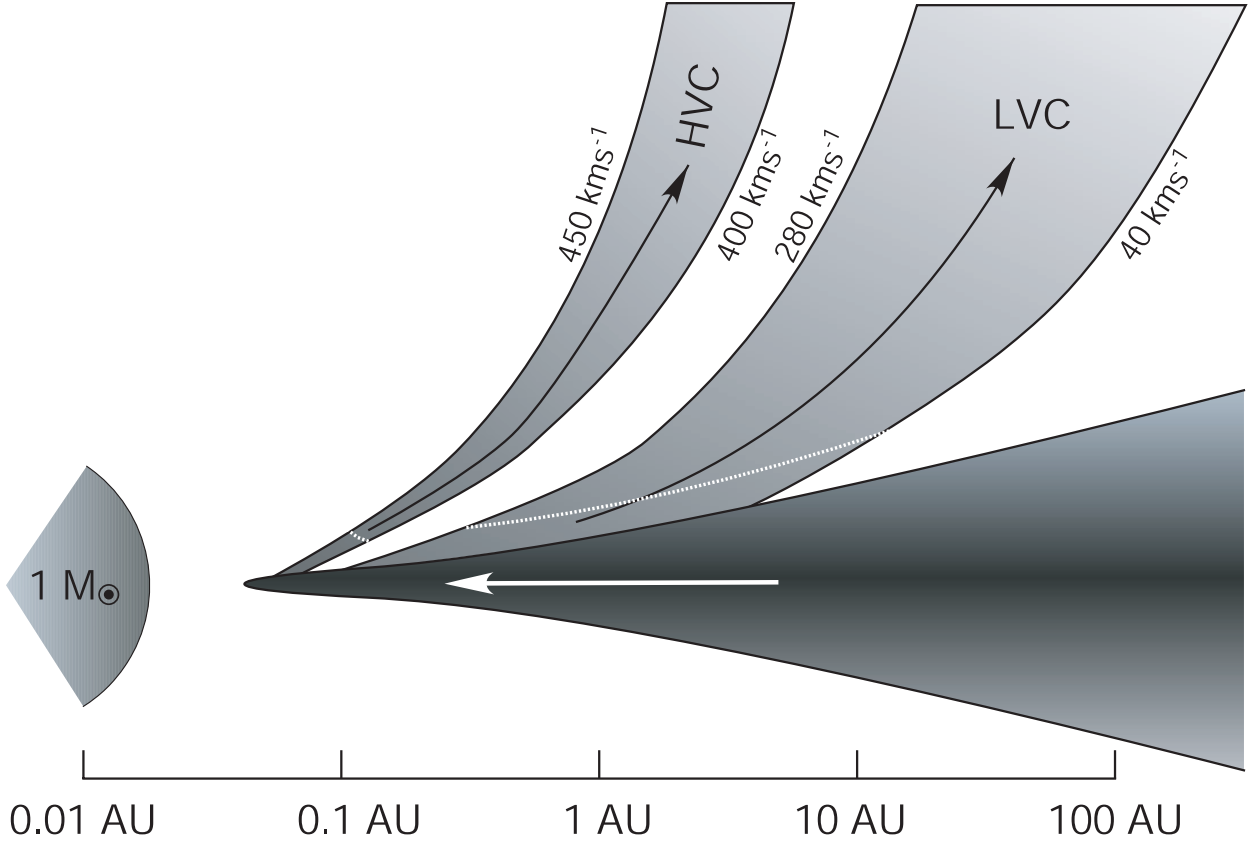}
\caption{
A schematic diagram showing the ranges of launching radius for a magnetocentrifugally accelerated outflows from a Keplerian-rotating accretion disk.
The central star is assumed to have a mass of 1 M$_{\odot}$.
The Alfv\'en radius, the locus of which is shown in white dotted-lines, is simply assumed to be three times that of the launching point radius anywhere on the disk.
The line of sight is 45$^{\circ}$ with respect to the polar axis, and the observed radial velocities ($V_{\rm rad}$) are $1/\sqrt{2}$ of the actual velocities ($V$).
A jet with $V_{\rm rad}$=280--320 \kms\ ($V$=400--450 \kms) is launched in a radial range of 0.04 to 0.05 AU, while a jet with $V_{\rm rad}$=30--200 \kms\  ($V$=40--280 \kms) is launched in a radial range of 0.1 to 4.5 AU.
The white arrow indicates the direction of the mass accretion in the disk.
\label{Schematic}}
\end{figure}

In summary, the presence of spatially extended HVCs and LVCs with very different kinematical characteristics suggests two types of outflows from a single driving source: the HVC is launched from an inner, narrow region of an accretion disk and the LVC from an outer, wider region.
Figure~\ref{Schematic} shows a schematic diagram showing the launching radii of the HVC and LVC for the northern jet of L1551 IRS~5.
We simply assumed a Keplerian-rotating accretion disk around a 1~M$_{\odot}$ star with the Alfv\'en radius three times that of the launching point radius along each magnetic line of force.
The HVC with a narrow velocity width is launched in a very small radial range from 0.04 to 0.05~AU from the star, while the LVC with a broader velocity width is launched in a radial range from 0.1 to 4.5~AU, 400 times larger radial range than that of the HVC.
The regions responsible for the HVC and LVC are spatially separated on the accretion disk. 
If both the HVC and LVC arise from a single outflow launched continuously over a radial range from 0.04 to 4.5~AU, we must seek physical and chemical conditions on the disk surface to heat up the gas along the two separated groups of streamlines.
Instead, if we assume two mechanisms working simultaneously--for example, if the HVC is driven by an $X$-wind type mechanism and the LVC is driven by a disk-wind type mechanism--it may be possible to have the two emitting gases along the two independent streamline groups as shown in Figure~\ref{Schematic}.

One of the interesting results of our current observations is shown in Figure~\ref{ChannelMaps}, where the LVC itself shows a variation in spatial width that varies with radial velocity, as quantitatively measured and listed in Table~\ref{jet_width}.
The spatial width of the LVC is $\sim$0\farcs65 in its very-blueshifted velocity channels (Figures~\ref{ChannelMaps}$g$ and \ref{ChannelMaps}$h$), while it is $\sim$0\farcs85 at its less-blueshifted velocity channels, shown in Figures~\ref{ChannelMaps}$m$ and \ref{ChannelMaps}$n$.
The difference in the spatial width of the LVC with velocity provides observational evidence that slower-velocity gas flows along outer and more widely-opened streamlines.
This may support the idea that the LVC (but not the HVC) of the L1551 IRS~5 northern jet is an outflow launched over a wide radial range of an accretion disk. Alternatively, part of the LVC-emitting gas flowing in the vicinity of the HVC may arise in the shocked gas entrained by the HVC.

The similar spatial width of the HVC compared to that of the very-blueshifted part ($V_{\rm LSR}\sim-$200~\kms) of the LVC may imply that the gas responsible for this part of the LVC flows immediately outside the HVC with a velocity difference of $\sim$100 \kms. The two flows may interact to produce the fainter ``pedestal'' emission connecting the HVC and LVC in the PVDs.
We discussed in Paper~I a possibility that the pedestal emission and the ``red wing emission'' at $Y\ge\,$2\farcs5 are produced as a result of turbulent entrainment of the LVC gas.
The pedestal emission has a velocity ($V_{\rm LSR}\sim-$200~\kms) similar to the secondary peak velocity ($V_{\rm LSR}\sim-$180~\kms) of the H$\alpha$ emission taken toward the base of the northern jet \citep{Stocke88} and the red wing emission shows a tendency of apparent acceleration with distance from IRS~5 VLA.
These characteristics favor the idea of pedestal and red wing emissions arising in the gas entrained by the HVC flow.
On the other hand, the main part of the LVC does not exhibit such apparent acceleration along the flow, although it does show velocity narrowing (Figure~\ref{VX_PVDs}), and is not affected by the time variation that was observed in the HVC at 0$''< Y <$1$''$.
This suggests that it is difficult to consider the entire LVC emission as arising from the entrained gas.

We also pointed out in Paper~II that the spatially wide subcomponent of the LVC (LVC$_{\rm WIDE}$) cannot be considered as arising from a bow shock because there is no systematic velocity difference or gradient between the LVC and LVC$_{\rm WIDE}$. 
On the other hand, the H$_2$ v=1--0 S(1) emission is less blueshifted ($V_{\rm LSR}\,$\rlap{$_\sim$}{$^<$}\,60~\kms) than \FeII\ and has a steadily increasing velocity component along the flow \citep{davis01}, suggesting that it originates from the gas entrained by the LVC or LVC$_{\rm WIDE}$.
Another possibility that the significant part of the LVC flow is a shock-decelerated HVC gas will be discussed below in relation to the x-ray emission. 

Finally, we should note that there is a third object recently discovered at 0\farcs09 southeast of the northern binary component \citep{lim06}.
The possibility that this third object contributes to either of the HVC or LVC is small because there is no hard evidence of an associated radio jet; the HVC and LVC of the L1551 IRS~5 northern jet are driven by a single source.
The HVC and LVC show exactly the same PAs as seen in Figure~\ref{ChannelMaps} and both are spatially confined around the $Y$-axis ($X=$0$''$) even at $Y\sim$4$''$ (see Figure~3$a$ of Paper~II). 
This level of alignment between the HVC and LVC is rather unlikely if different driving sources are  responsible for the HVC and LVC.
The peak of the HVC emission is systematically shifted to the positive $X$ direction by 0\farcs1--0\farcs2 with respect to that of the LVC at each position of $Y$. The reason for this is unknown, but such a systematical shift may also be unnatural if the HVC and LVC originate from different driving sources.

\subsection{The X-ray Source}\label{sec:Xray}
The soft x-ray emission associated with L1551 IRS~5 was first detected by \citet{favata02} with {\it XMM-Newton}, although the spatial resolution was not sufficient to pinpoint the actual location of the x-ray source on the jets.
Subsequently, \citet{bally03} located an x-ray source at 0$\dotsec$5--1$\arcsec$ west of IRS~5 VLA from {\it Chandra} observations.
This raised the possibility that the x-ray emission comes from PHK1, and \citet{bally03} proposed several models on this basis.
Because the actual location of the x-ray source is still uncertain, we re-examined the location of the x-ray emission peak.

We confirm in this paper that the {\it H}-band continuum peak, which agrees with the {\it K}-band peak within an uncertainty of 0\farcs25, is located 0\farcs6 east-northeast of PHK1 and the distance did not change between the two observing epochs separated by four years.
On the other hand, \citet{bally03} reported that the published infrared position for IRS~5 is about 0$\dotsec$5 or more east of the x-ray source.
These two measurements suggest that the peak position of the x-ray emission coincides with PHK1 within an uncertainty of $\sim$0\farcs3.
Note that this result does not rely on the radio position of IRS~5 VLA, which has a larger positional uncertainty as a result of a larger registration error between the optical and radio reference frames.

\citet{favata06} reported (based on their new {\it Chandra} observations) that a stronger, point-like component of the x-ray source did not show any detectable proper motion, while a weaker component expanded in size by $\sim$300~AU over four years, which corresponds to the projected velocity of $\sim$330~kms$^{-1}$.
It is interesting to compare these facts with another set of findings showing that the proper motion of PHK1 was not detected between the two epochs of observations in 2000 and 2004, while the HVC of the northern jet has a radial velocity of $\sim$300~kms$^{-1}$, which is equal to the projected velocity if we assume a jet inclination angle of 45$\arcdeg$.
Such similarities strengthen the possibility that the point-like x-ray emission originates in PHK1 and further implies that the HVC, but not the LVC, is responsible (at least) for the extended x-ray component. 
The extended and rapidly expanding x-ray component might be related to the newly seen knot at $Y$=0\farcs6 in Figure~\ref{FeII}.

The mechanism of the x-ray emission near PHK1 is still unclear.
The radial velocity of the HVC (260--350 \kms) corresponds to the actual velocity of 370--500 \kms\ with the above assumed inclination. 
Such an outflow has sufficiently high energy to produce a plasma of 10$^6$--10$^7$ K responsible for the x-ray emission when thermalized in a fast shock \citep{bally03,Raga02}.
Because the highly collimated HVC is present at PHK1 and in the region farther away from the driving source, it may be difficult to assume that the HVC is fully thermalized at PHK1.
However, the spatial coincidence of the HVC with the LVC (with its large velocity dispersion of 240 \kms\ around PHK1) might imply that part of the HVC is thermalized to produce the x-ray emission and LVC.
In such a case, a significant part of the LVC might be shock-decelerated HVC gas. 

\section{SUMMARY}
The main results of this paper are summarized as follows:
\begin {enumerate}
\item{
The L1551 IRS~5 VLA sources and the {\it H}-band continuum peak are located 1\farcs2 and 0\farcs6, respectively, east-northeast of the strongest [\ion{Fe}{2}] knot PHK1 along the northern jet.
The relative position of PHK1 with respect to the {\it H}-band continuum peak did not show any variation between the two observing epochs in 2000 and 2004, suggesting that PHK1 is stationary.
}
\item{
The southern jet significantly contributes to the \FeII\ flux at PHK1. 
It has a constant velocity of $\VLSR$=$-$120 \kms\ with a comparatively narrow velocity width of 53 \kms\ (deconvolved FWHM), and is traced as close as 0\farcs6 from the IRS~5 VLA sources.
}
\item{
The high- and low-velocity components (HVC and LVC) of the northern jet are clearly separated in position-velocity diagrams.
The LVC shows a wide velocity coverage from $-$240 to 0 \kms\ with a deconvolved FWHM width of 150--180 \kms.
It has a spatial width of 0\farcs6--0\farcs7, similar to that of the HVC, at its more-blueshifted velocity ($\sim-$200 \kms), while it shows a wider spatial width of 0\farcs8--0\farcs9 at its less-blueshifted velocity ($\sim-$30 \kms)
}
\item{
The high velocity and narrow line width of the HVC implies that its launching point is located over a narrow radial range in the inner part of an accretion disk, while the wide line width of the LVC means that it is launched from a wider radial range if it is also launched from the accretion disk.
The variation of the spatial width with velocity observed for the LVC suggests that slower gas flows in an outer annulus with a wider opening angle.
}
\item{
The velocity of the HVC shows a time variation between the two observing epochs in 2000 and 2004.
}
\item{
The position of the stationary, point-like x-ray source with respect to the near-infrared position of L1551 IRS~5 indicates that it coincides with PHK1 within an uncertainty of 0\farcs3. 
The projected expanding velocity of the weaker, extended component of the x-ray emission agrees well with the HVC velocity.
}

\end {enumerate}

\acknowledgments
We thank our referee, Dr. Rafael Bachiller, for providing us with valuable comments that have greatly improved this paper. 
We are grateful to the entire staff of the Subaru Telescope for their dedicated support to the telescope and observatory operations. 
This research has been made using NASA's Astrophysics Data System and the SIMBAD database operated at CDS, Strasbourg, France.


\clearpage
\begin{deluxetable}{cccccc}
\tablecolumns{6}
\tablewidth{0pt}
\tablecaption{Spatial width of the northern jet as a function of radial velocity\label{jet_width}} 
\tablehead{
\colhead{Channel} &\colhead{$V_{\rm LSR}$}&  \colhead{Peak Flux} & \colhead{Peak $X$} & \colhead{Peak $Y$} & \colhead{FWHM Width\tablenotemark{a}}\\
\colhead{in Figure~\ref{ChannelMaps}} &\colhead{(\kms)} &\colhead{($\times$ 10$^{-18}$ W m$^{-2}$ \AA$^{-1}$)} &\colhead{(arcsec)} &\colhead{(arcsec)} &\colhead{(arcsec)}
} 
\startdata 
$c$ & $-$355.6 to $-$326.1&0.13 & 0.18 & 1.20 & 0.58 \\
$d$ & $-$326.1 to $-$296.6 &0.23 & 0.12 & 0.90 & 0.56 \\
$e$ & $-$296.6 to $-$267.1 &0.17 & 0.06 & 0.60 & 0.68 \\
$f$ & $-$267.1 to $-$237.6 &0.12 & 0.06 & 0.12 & 0.77 \\
$g$ & $-$237.6 to $-$208.1 &0.12 & 0.18 & 0.84 & 0.64 \\
$h$ & $-$208.1 to $-$178.5 &0.17 & 0.12 & 0.84 & 0.66 \\
$i$ & $-$178.5 to $-$149.0 &0.22 & 0.00 &$-$0.06 & 0.96 \\
$j$ & $-$149.0 to $-$119.5 &0.38 &$-$0.12 & 0.12 & 1.05 \\
$k$ & $-$119.5 to $-$90.0 &0.44 &$-$0.18 & 0.00 & 0.97 \\
$l$ & $-$90.0 to $-$60.5 &0.38 & 0.00 &$-$0.06 & 0.85 \\
$m$ & $-$60.5 to $-$31.0 &0.30 &$-$0.06 &$-$0.24 & 0.86 \\
$n$ & $-$31.0 to $-$1.5 &0.13 & 0.06 &$-$0.42 & 0.83 \\
\enddata 
\tablenotetext{a}{Not deconvolved with the seeing size of 0\farcs4}
\end{deluxetable}

\clearpage
\begin{deluxetable}{lcccccl}
\tablecolumns{7}
\tablewidth{0pt}
\tablecaption{Deconvolved velocity widths (FWHM) of the HVCs and LVCs for \FeII\ Jets \label{velocity_width}} 
\tablehead{
\colhead{ } & \multicolumn{2}{c}{HVC} &\colhead{ }& \multicolumn{2}{c}{LVC}& \colhead{ } \\
\cline{2-3}\cline{5-6} 
\noalign{\vskip 2pt}
\hfil{Jet} & \colhead{Width} & \colhead{Remark} &\colhead{} & \colhead{Width} & \colhead{Remark} & \hfil{Reference} \\
\colhead{ }  & \colhead{(\kms)} &\colhead{ } &\colhead{ }&\colhead{(\kms)} & \colhead{} & \colhead{} 
} 

\startdata 
L1551 IRS~5 Northern Jet & 40 & Extended && 150--180 & Extended & Present work\\
DG Tau & 50 & Extended && $\sim$100 & Extended & \citet{pyo03}\\
RW Aur & 50 & Extended && 100 & Compact & \citet{pyo06}\\
HL Tau & 40 & Extended && $\ga$100 & Compact & \citet{pyo06}\\
L1551 IRS~5 Southern Jet & 53 & Extended && N/A & Not Detected & Present work\\
\enddata 
\end{deluxetable}

\clearpage

\clearpage

\clearpage

\clearpage

\clearpage

\clearpage

\end{document}